\documentstyle[12pt]{article}
\begin{document}

\tolerance=5000

\def\pp{{\, \mid \hskip -1.5mm =}}
\def\cL{{\cal L}}
\def\be{\begin{equation}}
\def\ee{\end{equation}}
\def\bea{\begin{eqnarray}}
\def\eea{\end{eqnarray}}
\def\tr{{\rm tr}\, }
\def\nn{\nonumber \\}
\def\e{{\rm e}}

\  \hfill
\begin{minipage}{3.5cm}
June 2003 \
\end{minipage}

\vfill

\begin{center}
{\large\bf Effective equation of state and energy conditions in 
phantom/tachyon inflationary cosmology perturbed by quantum effects.}

\vfill

{\sc Shin'ichi NOJIRI}\footnote{snojiri@yukawa.kyoto-u.ac.jp, nojiri@nda.ac.jp}
and {\sc Sergei D. ODINTSOV}$^{\spadesuit}
$\footnote{
odintsov@ieec.fcr.es
Also at Tomsk State Pedagogical University, Tomsk, RUSSIA.}

\vfill

{\sl Department of Applied Physics \\
National Defense Academy,
Hashirimizu Yokosuka 239-8686, JAPAN}

\vfill

{\sl $\spadesuit$ 
 Institut d'Estudis Espacials de Catalunya,
Consejo Superior de Investigaciones Cientificas (IEEC/CSIC)\\ 
Edifici Nexus, Gran Capit\`a 2-4, 08034, Barcelona, SPAIN
and Instituci\`o Catalana de Recerca i Estudis Avan\c{c}ats (ICREA)}


\begin{abstract}

We discuss the model consisting of tachyon which may have the negative
kinetic energy 
plus scalar phantom and plus conformal quantum matter. It is demonstrated 
that such a model naturally admits two deSitter phases where the early universe 
inflation is produced by quantum effects and the late time accelerating universe is caused 
by phantom/tachyon. The energy conditions bounds for such cosmology are derived.
It is interesting that effective equation of state may change its sign which depends
from the proper choice of the combination of phantom/tachyon and quantum effects.

\end{abstract}
\end{center}

\vfill


\newpage

\noindent
1. Recent astronomical data suggest the existence of the dark 
energy with negative pressure \cite{SuNv} which should provide
approximately two thirds of the current universe energy density.
It is expected that this dark energy is responsible for accelerating
expansion of the observable universe. 
The ratio $w$ between the pressure 
$p_d$ and the energy density $\rho_d$  
of the dark energy seems to be near or less than $-1$, $-1.62<w\equiv {p_d \over \rho_d}
< -0.74$ \cite{MMOT}. Numerous models of dark energy exist.
The phantom (field with negative kinetic energy) \cite{caldwell} was
also proposed as a candidate for dark energy as it admits negative
pressure.

 From another side, there is much interest now in the tachyon cosmology
(see \cite{tachyon} and for a review, \cite{GibbonsT}) where the
appearence of tachyon is basically motivated by string theory.
However, standard tachyon cosmology seems to be unsatisfactory 
and it is expected that stringy tachyon could be important in a possible 
pre-inflationary open string era.
Nevertheless, it is interesting that tachyon
with negative kinetic energy (another type of phantom) could be
introduced \cite{HL}. Such a phantom/tachyon  
model has naturally negative $w$. It admits the late time deSitter
attractor
solution and maybe considered as an interesting dark energy
model\cite{HL}. However, it is clear that universe should not be so simple 
and different matter should be present there. In particular,
it would be interesting to construct the accelerating universe 
with the inflationary early epoch (with possible positive or negative 
equation of
state) and with current accelerating universe dominated by dark energy.

The purpose of the present letter is aimed in this direction.
We consider a constituent model of tachyon  with potential
(tachyon may have the negative kinetic energy) plus standard phantom and 
quantum effects from the usual matter.
The effective equation of state is derived and discussed in detail. It is
shown that such a model
admits two deSitter phases where the first phase (early universe) is
produced by quantum effects and the late phase (current universe) is
produced by phantom/tachyon. The restrictions from energy conditions to
such cosmology are discussed.

\ 

\noindent
2. The starting tachyon action  is given by
\be
\label{TC1}
S_\phi = -\int d^4 \sqrt{-g}\left\{V(\phi)\sqrt{1 + \lambda g^{\mu\nu}\partial_\mu \phi
\partial_\nu \phi} + U(\phi)\right\}\ .
\ee
When $\lambda=1$ and $U(\phi)=0$, the above action describes the usual tachyon but 
when $\lambda = -1$ and $U(\phi)=0$ it corresponds to the phantom/tachyon
as in \cite{HL}.

The simplest way to account for quantum effects 
(at least, for CFT matter) is to include the contributions due to
conformal anomaly:
\be
\label{OVII}
T=b\left(F+{2 \over 3}\Box R\right) + b' G + b''\Box R\ ,
\ee
where $F$ is the square of 4d Weyl tensor, $G$ is Gauss-Bonnet invariant. 
In general, with $N$ scalar, $N_{1/2}$ spinor, $N_1$ vector fields, $N_2$  ($=0$ or $1$) 
gravitons and $N_{\rm HD}$ higher derivative conformal scalars, $b$, $b'$ and $b''$ are 
given by
\bea
\label{bs}
&& b={N +6N_{1/2}+12N_1 + 611 N_2 - 8N_{\rm HD} 
\over 120(4\pi)^2}\nn 
&& b'=-{N+11N_{1/2}+62N_1 + 1411 N_2 -28 N_{\rm HD} 
\over 360(4\pi)^2}\ , \quad b''=0\ .
\eea
Let the metric of the 4 dimensional spacetime has the warped form:
\be
\label{ddS1}
ds^2= - dt^2 + L^2\e^{2A}\sum_{i,j=1}^3\left(dx^i\right)^2\ .
\ee
Then the contributions due to conformal anomaly to $\rho$ and $p$ are 
found explicitly in \cite{NOev,NOOfrw}. 
The ``radius'' of the universe $a$ and the Hubble parameter $H$ are
introduced as follows
\be
\label{en6}
a\equiv L\e^A\ ,\quad H={1 \over a}{d a \over dt}
={d A \over dt}\ .
\ee
FRW equation looks like:
\bea
\label{TC2}
H^2 &=& {\kappa \over 3}\left(\rho_\phi+\rho_A\right)\ ,\\
\label{TC3}
{\ddot a \over a}&=&-{\kappa \over 3}\left\{{1 \over 2}\left(\rho_\phi + \rho_A\right) 
+ {3 \over 2}\left(p_\phi + p_A\right)\right\}\ .
\eea
Here $\rho_\phi$ and $p_\phi$ are the energy density and the pressure of
the tachyon $\phi$, respectively. 

The natural assumption is that $\phi$ only depends on time $t$. Then  
the equation of the motion for tachyon follows from 
(\ref{TC1}):
\be
\label{TC4}
0=\lambda\ddot \phi + \left(3\lambda H\dot\phi + {V'(\phi) \over V(\phi)}\right)
\left(1 - \lambda \dot\phi^2\right) + {U'(\phi) \over V(\phi)}
\left(1 - \lambda \dot\phi^2\right)^{3 \over 2}\ .
\ee 
The energy density $\rho_\phi$ and the pressure $p_\phi$ are given by
\be
\label{TC5}
\rho_\phi={V(\phi) \over \sqrt{1 - \lambda \dot\phi^2}}+ U(\phi)\ ,\quad 
p_\phi=-V(\phi) \sqrt{1 - \lambda \dot\phi^2} - U(\phi)\ .
\ee 
We may further assume that the spacetime is deSitter space: $a=\e^{t \over L}$. 
Then \cite{NOev,NOOfrw}:
\be
\label{phtm2}
\rho_A=-p_A = - {6b' \over L^4}\ .
\ee
and the FRW equations (\ref{TC2}) and (\ref{TC3}) are:
\bea 
\label{TC6}
{1 \over L^2}&=&{\kappa \over 3}\left(
{V(\phi) \over \sqrt{1 - \lambda \dot\phi^2}}+ U(\phi) - {6b' \over L^4}\right) \ ,\\
\label{TC7}
{1 \over L^2}&=&-{\kappa \over 3}\left({V(\phi) \over 2\sqrt{1 - \lambda \dot\phi^2}}
+ {3 \over 2}V(\phi) \sqrt{1 - \lambda \dot\phi^2} - U(\phi) + {6b' \over L^4}\right) \ .
\eea
By combining (\ref{TC6}) and (\ref{TC7}), we find $\dot\phi=0$, that is, 
$\phi$ should be a constant:
\be
\label{TC7b}
\phi=\phi_0 \ .
\ee
Then Eq.(\ref{TC6}) gives
\be
\label{TC8}
{1 \over L^2}={\kappa \over 3}\left(
V(\phi_0) + U(\phi_0) - {6b' \over L^4}\right) \ .
\ee
On the other hand, eq. of motion (\ref{TC4}) gives
\be
\label{TC9}
V'(\phi_0) + U'(\phi_0)=0\ .
\ee
Then $V(\phi_0) + U(\phi_0)$ has an extremum at $\phi=\phi_0$ or 
$V(\phi_0) + U(\phi_0)$ is a constant. The above aruguments 
are valid even if $U(\phi)=0$ if $V(\phi)$ has an extremum.  
If we follow the proposal in \cite{KMM}, $V(\phi)$ has a following form:
\be
\label{TC10}
V(\phi)=V_0 \left( 1 + {\phi \over \phi_0}\right)\e^{-{\phi \over \phi_0}}\ ,
\ee
which is monotonically decreasing function with respect $\phi$ and has  
extrema at $\phi=0$, where $V(\phi)=V_0$, and at $\phi=+\infty$, $V(\phi)=0$. 
If, for example, 
\be
\label{TC11}
U(\phi)= - V_0 {\phi \over \phi_0}\e^{-{\phi \over \phi_0}}\ ,
\ee
$V(\phi) + U(\phi)$ is always constant. 
Since $\dot\phi=0$, we find $\rho_\phi=-p_\phi=V(\phi_0) + U(\phi_0)$ and 
therefore $w=-1$. 
One can solve (\ref{TC8}) with respect to $L^2$:
\be
\label{TC12}
{1 \over L^2}=-{1 \over 4b'\kappa}\pm \sqrt{{1 \over 16 {b'}^2 \kappa^2} 
+ {U_0 \over 6b'}}\ .
\ee
Here 
\be
\label{TC13}
U_0\equiv V(\phi_0) + U(\phi_0)\ .
\ee
Since $b'$ is usually negative, if
\be
\label{TC14}
V_0\leq - {3 \over 8b'\kappa^2}\ ,
\ee
the both of the solutions are real and if $V_0>0$, both of solutions are positive. 
We may consider the case that $V_0$ is small (more exactly 
$\left| b'\kappa^2 V_0\right|\ll 1$), then 
\be
\label{TC15}
{1 \over L^2} \sim - {1 \over 2b'\kappa},\  {\kappa U_0 \over 3}\ .
\ee
This suggests a senario where the early universe expands by 
${1 \over L^2} \sim - {1 \over 2b'\kappa}$ (anomaly driven inflation) and
the late universe (at present) 
expands by the smaller solution ${1 \over L^2} \sim {\kappa U_0 \over 3}$. 
Eq.(\ref{TC15}) shows that there are two regimes, one is given by almost
purely 
quantum effects (slightly perturbed by tachyon) and another one is due to almost only 
tachyon perturbed by quantum effects. Then quantum induced inflation and tachyon induced one 
effectively decouple in such scenario.

Since now $\dot\phi=0$, one gets
\be
\label{TC16}
w={p_\phi \over \rho_\phi}={p_A \over \rho_A}={p_\phi + p_A \over \rho_\phi + \rho_A}=-1\ .
\ee
For more general case with $\dot\phi\neq 0$ in (\ref{TC5}), the effective
equation of state is
\be
\label{TC17}
w_\phi\equiv {p_\phi \over \rho_\phi}=-1 
+ {\lambda \dot\phi^2 \left( V(\phi) - {\lambda\dot\phi^2 U(\phi) 
\over \sqrt{1 - \lambda \dot\phi^2}} \right) \over 
V(\phi) + U(\phi) \sqrt{1 - \lambda \dot\phi^2}}\ .
\ee
Therefore when $V(\phi)>0$ and $U(\phi)\geq 0$, if $\lambda<0$, that is, 
the tachyon is phantom-like, we find $w\leq -1$. 

\ 

\noindent
3.We may further couple the above model with the usual phantom field
\cite{GibbonsP} (deSitter universe induced by phantom with quantum effects 
was studied in \cite{NOp}). 
The energy density $\rho_C$ and pressure $p_C$ of the 
phantom field is given by \cite{GibbonsP}
\be
\label{TC18}
\rho_C = p_C = - {C^2 \over 2}\ ,
\ee
where $C$ is a constant. Denoting the energy density and pressure of the 
matter by $\rho_m$ and $p_m$, respectively, Eqs.(\ref{TC6}) and (\ref{TC7}) are 
modified as
\bea
\label{TC19}
{1 \over L^2}&=&{\kappa \over 3}\left(
{V(\phi) \over \sqrt{1 - \lambda \dot\phi^2}}+ U(\phi) - {6b' \over L^4}
 - {C^2 \over 2} + \rho_m\right) \ ,\\
\label{TC20}
{1 \over L^2}&=&-{\kappa \over 3}\left({V(\phi) \over 2\sqrt{1 - \lambda \dot\phi^2}}
+ {3 \over 2}V(\phi) \sqrt{1 - \lambda \dot\phi^2} - U(\phi) + {6b' \over L^4} \right. \nn
&& \left. \qquad - C^2 + {1 \over 2}\rho_m + {3 \over 2}p_m\right) \ .
\eea
Eqs.(\ref{TC19}) and (\ref{TC20}) can be solved with respect to $\rho_m$ and $p_m$: 
\bea
\label{TC21}
\rho_m&=& - {V(\phi) \over \sqrt{1 - \lambda \dot\phi^2}} - U(\phi) + {6b' \over L^4}
+ {C^2 \over 2} + {3 \over \kappa L^2}\ ,\\
\label{TC22}
p_m&=& V(\phi) \sqrt{1 - \lambda \dot\phi^2} + U(\phi) - {6b' \over L^4}
+ {C^2 \over 2} - {3 \over \kappa L^2}\ .
\eea

Let us remind  the standard energy conditions accepted in cosmology:
\bea
\label{phtm11}
&\circ &\ \mbox{Null Energy Condition (NEC):} \ \rho + p \geq 0\\
\label{phtm8}
&\circ &\ \mbox{Weak Energy Condition (WEC):} \ \rho\geq 0 \ \mbox{and}\ \rho + p \geq 0 \\
\label{phtm9}
&\circ &\ \mbox{Strong Energy Condition (SEC):} \ \rho + 3 p \geq 0\ \mbox{and}\ \rho + p \geq 0\\
\label{phtm10}
&\circ &\ \mbox{Dominant Energy Condition (DEC):}\ \rho\geq 0 \ \mbox{and}\ \rho \pm p \geq 0 
\eea
It is interesting to analyze the restrictions to current deSitter cosmology from
the energy conditions.\footnote{Generally speaking, the energy conditions 
bounds are not dictated by some deep physical principle.}
By combining (\ref{TC21}) and (\ref{TC22}), one gets
\be
\label{TC23}
\rho_m + p_m = C^2 - {\lambda\dot\phi^2 V(\phi) \over \sqrt{1-\lambda\dot\phi^2}}\ .
\ee
Then NEC can be satisfied if $V(\phi)>0$, $\lambda<0$, which is the case that the 
tachyon is phantom-like. The expression of $\rho_m$  (\ref{TC21}) can be 
rewritten as
\bea
\label{TC24}
\rho_m &=& {\beta(\phi) \over 2L^4}\left( L^4 - {6 \over \kappa \beta(\phi)}L^2  
+ {12b' \over \beta(\phi)}\right) \nn
&=& {\beta(\phi) \over 2L^4}\left( L^2 - {3 \over \kappa \beta(\phi)} 
+ \sqrt{\left({3 \over \kappa \beta(\phi)}\right)^2 - {12b' \over \beta(\phi)}}\right) \nn
&& \times \left( L^2 - {3 \over \kappa \beta(\phi)} - \sqrt{\left({3 \over \kappa \beta(\phi)}
\right)^2 - {12b' \over \beta(\phi)}}\right)\ , \\
\beta(\phi)&\equiv& {C^2 \over 2} - { V(\phi) \over \sqrt{1-\lambda\dot\phi^2}} 
 - U(\phi)\ .\nonumber
\eea
If $\beta(\phi)<0$, $\rho_m<0$, then WEC or DEC is not satisfied. If $\beta(\phi)>0$, 
since $L^2 - {3 \over \kappa \beta(\phi)} 
+ \sqrt{\left({3 \over \kappa \beta(\phi)}\right)^2 - {12b' \over \beta(\phi)}}>0$, 
WEC or DEC gives a non-trivial constraint on $L^2$:
\be
\label{TC25}
L^2 > {3 \over \kappa \beta(\phi)} + \sqrt{\left({3 \over \kappa \beta(\phi)}
\right)^2 - {12b' \over \beta(\phi)}}\ .
\ee
We also have
\bea
\label{TC26}
\rho_m + 3p_m &=& {\gamma(\phi) \over L^4}\left( L^4 - {6 \over \kappa \gamma(\phi)}L^2  
 - {12b' \over \gamma(\phi)}\right) \nn
&=& {\gamma(\phi) \over L^4}\left( L^2 - {3 \over \kappa \gamma(\phi)} 
+ \sqrt{\left({3 \over \kappa \gamma(\phi)}\right)^2 + {12b' \over \gamma(\phi)}}\right) \nn
&& \times \left( L^2 - {3 \over \kappa \gamma(\phi)} - \sqrt{\left({3 \over \kappa \gamma(\phi)}
\right)^2 + {12b' \over \gamma(\phi)}}\right)\ , \\
\gamma(\phi)&\equiv& 2C^2 - { V(\phi) \over \sqrt{1-\lambda\dot\phi^2}}
+ 3 V(\phi) \sqrt{1-\lambda\dot\phi^2} + 2 U(\phi)\ .\nonumber
\eea
When $V(\phi), U(\phi)>0$ and $\lambda<0$, we find $\gamma(\phi)>0$. 
In this case, if the quantity inside the squre root is negative:
\be
\label{TC27}
\left({3 \over \kappa \gamma(\phi)}\right)^2 + {12b' \over \gamma(\phi)}<0\ ,
\ee
we obtain $\rho_m + 3p_m > 0$ and SEC is satisfied. On the other hand, if
\be
\label{TC28}
\left({3 \over \kappa \gamma(\phi)}\right)^2 + {12b' \over \gamma(\phi)}>0\ ,
\ee
SEC gives a non-trivial constraint on $L^2$
\be
\label{TC29}
L^2 < {3 \over \kappa \gamma(\phi)} - \sqrt{\left({3 \over 
\kappa \gamma(\phi)}\right)^2 + {12b' \over \gamma(\phi)}}\ \mbox{or}\ 
L^2 > {3 \over \kappa \gamma(\phi)} + \sqrt{\left({3 \over 
\kappa \gamma(\phi)}\right)^2 + {12b' \over \gamma(\phi)}}\ .
\ee
One also has
\bea
\label{TC30}
\rho_m - p_m &=& -{\eta(\phi) \over L^4}\left( L^4 - {6 \over \kappa \eta(\phi)}L^2  
 - {12b' \over \eta(\phi)}\right) \nn
&=& -{\eta(\phi) \over L^4}\left( L^2 - {3 \over \kappa \eta(\phi)} 
+ \sqrt{\left({3 \over \kappa \eta(\phi)}\right)^2 + {12b' \over \eta(\phi)}}\right) \nn
&& \times \left( L^2 - {3 \over \kappa \eta(\phi)} - \sqrt{\left({3 \over \kappa \eta(\phi)}
\right)^2 + {12b' \over \eta(\phi)}}\right)\ , \\
\eta(\phi)&\equiv& {V(\phi) \over \sqrt{1-\lambda\dot\phi^2}}
+ V(\phi) \sqrt{1-\lambda\dot\phi^2} + 2 U(\phi)\ .\nonumber
\eea
Now if $V(\phi), U(\phi)>0$, then $\eta(\phi)>0$ and DEC gives a 
constraint:
\be
\label{TC31}
{3 \over \kappa \eta(\phi)} - \sqrt{\left({3 \over \kappa \eta(\phi)}
\right)^2 + {12b' \over \eta(\phi)}}< L^2 <
{3 \over \kappa \eta(\phi)} + \sqrt{\left({3 \over \kappa \eta(\phi)}
\right)^2 + {12b' \over \eta(\phi)}}\ .
\ee

As a special case, we consider the dust as a matter, where $p_m=0$. 
Then Eq.(\ref{TC22}) gives
\bea
\label{TC32}
 L^2 &=& {3 \over \kappa \xi(\phi)} \pm \sqrt{\left({3 \over \kappa \xi(\phi)}\right)^2 
+ {12b' \over \xi(\phi)}}\ , \\
\xi(\phi)&\equiv& {C^2 \over 2} + V(\phi) \sqrt{1-\lambda\dot\phi^2} + U(\phi)\ .\nonumber
\eea
Note that $\xi(\phi)$ is positive if $V(\phi), U(\phi)>0$. On the other hand, 
say, from (\ref{TC23}), we obtain
\be
\label{TC33}
\rho_m = C^2 - {\lambda\dot\phi^2 V(\phi) \over \sqrt{1-\lambda\dot\phi^2}}\ .
\ee
Then if $V(\phi)>0$, $\lambda<0$, which is the case that the 
tachyon is phantom-like, all the energy conditions are satisfied.  

When 
\be
\label{TC33b}
\left| \kappa^2 b'\xi(\phi)\right|\ll 1\ ,
\ee
 two solutions (\ref{TC32}) are:
\be
\label{TC33c} 
L^2 = {6 \over \kappa \xi(\phi)}\ ,\quad 
-2b' \kappa^2\ .
\ee 
The first one corresponds to the classical tachyon solution. On the other
hand, the second 
solution is induced by the quantum corrections and is independent of the
tachyon $\phi$ or 
phantom $C$. Then as in (\ref{TC15}), there are two decoupled 
deSitter phases. One is induced by quantum effects
and another one is  due to the classical tachyon and phantom. 

On the other hand, in (\ref{TC32}), we may consider the case that the
quantum contribution has almost same magnitude with that from the phantom
and tachyon: 
\be
\label{TC33d}
\left| \kappa^2 b'\xi(\phi)\right|\sim 1\ ,
\ee
Then  two solutions (\ref{TC32}) are of the same order with each
other.   
Especially when 
\be
\label{TC33e}
\kappa^2 b'\xi(\phi)= - 1\ ,
\ee
the solutions become degenerate and there is only one solution. The
magnitude of $L^2$ 
in the degenerate case is half of the classical case that $b'=0$. If 
\be
\label{TC33f}
\kappa^2 b'\xi(\phi)< - 1\ ,
\ee
the solutions (\ref{TC32}) become imaginary. This is indication to
(non-physical) oscillating metric. 

Let us investigate the effective equation of state, i.e. $w$ including all
the dark-side: tachyon, phantom and 
quantum corrections. By combining (\ref{TC5}), (\ref{phtm2}), and
(\ref{TC18}), 
one finds
\bea
\label{TC34}
w&\equiv&{p_\phi + p_A + p_C \over \rho_\phi + \rho_A + \rho_C} \nn
&=&{-V(\phi) \sqrt{1 - \lambda \dot\phi^2} - U(\phi)
+ {6b' \over L^4} - {C^2 \over 2} \over 
{V(\phi) \over \sqrt{1 - \lambda \dot\phi^2}}+ U(\phi) - {6b' \over L^4}- {C^2 \over 2}} \nn
&=& 1 - {V(\phi) \left(\sqrt{1 - \lambda \dot\phi^2} 
+ {1 \over \sqrt{1 - \lambda \dot\phi^2}}\right) + U(\phi) - {6b' \over L^4} \over 
{V(\phi) \over \sqrt{1 - \lambda \dot\phi^2}}+ U(\phi) - {6b' \over L^4}- {C^2 \over 2}} \ .
\eea
In the limit of $C\to\infty$,  $w\to 1$. On the other hand, when $C\to 0$, we find 
\bea
\label{TC34b}
w&\to &{-V(\phi) \sqrt{1 - \lambda \dot\phi^2} - U(\phi)
+ {6b' \over L^4} \over {V(\phi) \over \sqrt{1 - \lambda \dot\phi^2}}
+ U(\phi) - {6b' \over L^4}} \nn
&=& -1 + {\lambda \dot\phi^2 V(\phi) \over \sqrt{1 - \lambda \dot\phi^2}\left( 
{V(\phi) \over \sqrt{1 - \lambda \dot\phi^2}}+ U(\phi) - {6b' \over L^4}\right)} \ ,
\eea
which is less than $-1$ if $\lambda<0$, $U,V>0$, and $b'<0$. The obtained expression 
 (\ref{TC34b}) is a monotonically decreasing function of $C$ in the region 
$0<{C^2 \over 2} <{V(\phi) \over \sqrt{1 - \lambda \dot\phi^2}}+ U(\phi) - {6b' \over L^4}$ 
and ${V(\phi) \over \sqrt{1 - \lambda \dot\phi^2}} + U(\phi) - {6b' \over L^4}<
{C^2 \over 2} <\infty$. 
There is a pole singularity at ${C^2 \over 2} = {V(\phi) \over 
\sqrt{1 - \lambda \dot\phi^2}} + U(\phi) - {6b' \over L^4}$. In the limit 
${C^2 \over 2} \to {V(\phi) \over \sqrt{1 - \lambda 
\dot\phi^2}}+ U(\phi) - {6b' \over L^4} - 0$, we obtain $w\to -\infty$. Then in the 
region ${C^2 \over 2} \sim {V(\phi) \over \sqrt{1 - \lambda \dot\phi^2}}
+ U(\phi) - {6b' \over L^4}$ but ${C^2 \over 2} <{V(\phi) \over 
\sqrt{1 - \lambda \dot\phi^2}}+ U(\phi) - {6b' \over L^4}$, one may get large 
negative $w$. When ${C^2 \over 2} \sim {V(\phi) \over \sqrt{1 - \lambda \dot\phi^2}}
+ U(\phi) - {6b' \over L^4}$, from Eqs.(\ref{TC21}) and (\ref{TC22}), we
find
\bea
\label{TC35}
\rho_m&\sim & {3 \over \kappa L^2}\ ,\\
\label{TC36}
p_m&\sim& V(\phi) \left(\sqrt{1 - \lambda \dot\phi^2} + {1  \over 
\sqrt{1 - \lambda \dot\phi^2}}\right) + 2U(\phi) - {12b' \over L^4} - {3 \over \kappa L^2}\ .
\eea
Then NEC (\ref{phtm11}) and WEC (\ref{phtm8}) are satisfied. 
SEC  (\ref{phtm9}) is satisfied if 
\be
\label{TC37}
3V(\phi) \left(\sqrt{1 - \lambda \dot\phi^2} + {1  \over 
\sqrt{1 - \lambda \dot\phi^2}}\right) + 6U(\phi) - {36b' \over L^4} \geq {6 \over \kappa L^2}\ .
\ee
and DEC (\ref{phtm10}) is satisfied if 
\be
\label{TC38}
V(\phi) \left(\sqrt{1 - \lambda \dot\phi^2} + {1  \over 
\sqrt{1 - \lambda \dot\phi^2}}\right) + 2U(\phi) - {12b' \over L^4} \leq {6 \over \kappa L^2}\ .
\ee
Eqs.(\ref{TC37}) and (\ref{TC38}) show that SEC and DEC do not conflict
with each 
other. Although the value of $L^2$ depends on the contribution of the matter energy 
density, Eqs.(\ref{TC37}) and (\ref{TC38}) give non-trivial constraints on $L^2$. 
If
\be
\label{TC40}
\zeta(\phi)\equiv V(\phi) \left(\sqrt{1 - \lambda \dot\phi^2} + {1  \over 
\sqrt{1 - \lambda \dot\phi^2}}\right) + 2U(\phi) > 0\ ,
\ee
Eqs.(\ref{TC37}) and (\ref{TC38}) can be rewritten as
\be
\label{TC41}
L^2 \leq {1 \over \kappa^2 \zeta(\phi)} - \sqrt{{1 \over \kappa^4 \zeta(\phi)^2} 
+ {12b' \over \zeta(\phi)}} \quad \mbox{or}
L^2 \geq {1 \over \kappa^2 \zeta(\phi)} + \sqrt{{1 \over \kappa^4 \zeta(\phi)^2} 
+ {12b' \over \zeta(\phi)}}\ .
\ee
On the other hand, Eq.(\ref{TC38}) can be rewritten as
\be
\label{TC41b}
{3 \over \kappa^2 \zeta(\phi)} - \sqrt{{9 \over \kappa^4 \zeta(\phi)^2} 
+ {12b' \over \zeta(\phi)}} \leq
L^2 \leq {3 \over \kappa^2 \zeta(\phi)} + \sqrt{{9 \over \kappa^4 \zeta(\phi)^2} 
+ {12b' \over \zeta(\phi)}}\ .
\ee
Then all the energy conditions can be satisfied if
\bea
\label{TC41c}
&& {3 \over \kappa^2 \zeta(\phi)} - \sqrt{{9 \over \kappa^4 \zeta(\phi)^2} 
+ {12b' \over \zeta(\phi)}} \leq
L^2 \leq {1 \over \kappa^2 \zeta(\phi)} - \sqrt{{1 \over \kappa^4 \zeta(\phi)^2} 
+ {12b' \over \zeta(\phi)}} \nn 
\mbox{or} &&
{1 \over \kappa^2 \zeta(\phi)} + \sqrt{{1 \over \kappa^4 \zeta(\phi)^2} 
+ {12b' \over \zeta(\phi)}} \leq L^2 \leq {3 \over \kappa^2 \zeta(\phi)} 
+ \sqrt{{9 \over \kappa^4 \zeta(\phi)^2} + {12b' \over \zeta(\phi)}}\ .
\eea

We also note that in the limit ${C^2 \over 2} \to {V(\phi) \over \sqrt{1 - \lambda 
\dot\phi^2}}+ U(\phi) - {6b' \over L^4} + 0$ (instead of ${C^2 \over 2} \to {V(\phi) 
\over \sqrt{1 - \lambda \dot\phi^2}}+ U(\phi) - {6b' \over L^4} - 0$), we obtain 
$w\to +\infty$. Then when the contribution of the phantom has almost
the same magnitude with those 
from the tachyon and the quantum effects but that of the phantom is a
little bit bigger, 
$w$ becomes positive. Note that the sign of the contribution to the pressure from 
the phantom is opposite to those from the tachyon and the quantum
corrections. Then the 
sign change of $w$ occurs when the sign of the pressure is changed. 
We should also note that $w$ corresponding to only tachyon is given by (\ref{TC17}), 
which is less than or equal to $-1$ when $V(\phi)>0$, $U(\phi)\geq 0$, $\lambda<0$. 
On the other hand $w$ corresponding to quantum corrections is $-1$ as in (\ref{TC16}). 
Eq.(\ref{TC18}) tells $w$ corresponding to phantom is unity. 
The reason why $w$ can be positive when ${C^2 \over 2} > {V(\phi) \over 
\sqrt{1 - \lambda \dot\phi^2}}+ U(\phi) - {6b' \over L^4}$ but 
${C^2 \over 2} \sim {V(\phi) \over \sqrt{1 - \lambda  \dot\phi^2}}
+ U(\phi) - {6b' \over L^4} + 0$ is surely due to the existence of the
phantom. Moreover, 
 $w$ can become large due to the proper choice of  combination for
phantom, tachyon 
and the quantum effects. 

When $\dot \phi$ is small, the role of quantum effects is just to shift
the contribution from the tachyon:
\be
\label{TC39}
V(\phi) \sqrt{1 - \lambda \dot\phi^2} + U(\phi)
\sim {V(\phi) \over \sqrt{1 - \lambda \dot\phi^2}} + U(\phi)
\sim V(\phi) + U(\phi) \to V(\phi) + U(\phi)  - {6b' \over L^4}\ .
\ee
The quantum effects become dominant for small $L$ even if $C$
is large (compare with \cite{NOp}).
In the limit where $L$ is small, $w$ (\ref{TC34b}) goes to $-1$. 

The main lesson drawn from this study is that constituent dark energy
model may provide several deSitter-like phases of accelerated expansion 
where effective equation of state may change its sign when fine-tuning the
model. The role of quantum effects is to drive the early universe to
inflationary epoch. In some situations,
bounds provided by the standard energy conditions maybe satisfied.

\ 

\noindent
{\bf Acknowledgments.} The research is supported in part by the Ministry of
Education, Science, Sports and Culture of Japan under the grant n.13135208
(S.N.), DGI/SGPI (Spain) project BFM2000-0810 (S.D.O.), RFBR grant
03-01-00105
(S.D.O.) and LRSS grant 1252.2003.2
(S.D.O.).

\end{document}